# Efficient construction of effective Hamiltonians with a hybrid machine learning method

Yang Cheng[1*], Binhua Zhang[1*], Xueyang Li[1], Hongyu Yu[1],

Changsong Xu[1,2] and Hongjun Xiang[1†]

[1]*Key Laboratory of Computational Physical Sciences (Ministry of Education),*

*Institute of Computational Physical Sciences,*

*State Key Laboratory of Surface Physics, and Department of Physics, Fudan*

*University, Shanghai 200433, China*

[2]*Hefei National Laboratory, Hefei 230088, China*

The effective Hamiltonian method is a powerful tool for simulating large-scale systems across a wide range of temperatures. However, previous methods for constructing effective Hamiltonian models suffer from key limitations: some require to manually predefine interaction terms limited flexibility in capturing complex systems, while others lack efficiency in selecting optimal interactions. In this work, we introduce the Lasso-GA Hybrid Method (LGHM), a novel approach that combines Lasso regression and genetic algorithms to rapidly construct effective Hamiltonian models. Such method is broadly applicable to both magnetic systems (e.g., spin Hamiltonians) and atomic displacement models. To verify the reliability and usefulness of LGHM, we take monolayer $CrI_3$ and $Fe_3GaTe_2$ as examples. In both cases, LGHM not only successfully identifies key interaction terms with high fitting accuracy, but also reproduces experimental magnetic ground states and Curie temperatures with further Monte Carlo simulations. Notable, our analysis of monolayer $Fe_3GaTe_2$ reveals that the single-ion anisotropy and Heisenberg interaction lead to an out-of-plane ferromagnetic ground state, while the fourth-order interactions contribute significantly to the high Curie temperature. Our method is general so it can be applied to construct other effective Hamiltonian models.

*These authors contributed equally to this work.
†Contact author: hxiang@fudan.edu.cn

First-principle calculation has emerged as a crucial tool for studying the structural and electronic properties of solid-state materials. [1,2] Despite its utility, traditional Density Functional Theory (DFT) methods face significant challenges due to their high computational cost, making it difficult to explore finite-temperature properties or handle large-scale systems. Moreover, conventional DFT calculations typically yield ground state properties and final results without directly elucidating the underlying microscopic mechanisms. To overcome these limitations, effective Hamiltonian method, which is both computationally efficient and capable of delivering deeper physical insights, has gained widespread adoption for tackling thermodynamic and dynamic problems. The effective Hamiltonian method has been applied to investigate various different physical systems including ferroelectric systems and spin systems. [3–5] For instance, effective Hamiltonians of spin models provide insights not only into single-ion anisotropy (SIA), exchange coupling, and Dzyaloshinskii–Moriya interaction (DMI) [6] but also into all possible interaction forms permitted by the system's symmetry.

Our work focuses on deriving the effective Hamiltonian from the DFT calculation, which primarily involves constructing a Hamiltonian model and fitting it to the DFT outcomes. [3,4,7] This process determines the coefficients of various interaction terms, enabling further investigations and analyses. To demonstrate this workflow, let us take the effective

Hamiltonian model for magnetic systems as an example. When the interaction terms are truncated to two-body and second-order interactions, the spin component of the Hamiltonian describing the magnetic system can be expressed as: $H_{spin} = \sum_{i<j} J_{ij} \vec{S}_i \cdot \vec{S}_j + \sum_{i<j} \vec{D}_{ij} \cdot (\vec{S}_i \times \vec{S}_j) + \sum_i A_i S_{iz}^2$ , where the symmetric Heisenberg exchange interactions, antisymmetric Dzyaloshinskii–Moriya interactions and single-ion anisotropy are considered. Several methods have been developed to estimate the spin interaction parameters. [8–11] For example, the four-state method, where four specific states are DFT calculated to obtain the interaction parameters. This method is among the simplest and most efficient, and it has been extensively utilized in various magnetic systems. However, high-order interactions such as the biquadratic term [12] are sometimes important to the physical properties. For instance, the non-Heisenberg behavior in TbMnO$_3$ was successfully explained by the high-order term. When considering the high-order terms, the number of possible interaction terms often becomes large, and least-squares fitting is typically employed to estimate the parameters. [7,13,14] These methods, usually require pre-selecting which interaction terms to be included in the model before performing linear regression. In contrast, the machine learning based method [7] can automatically identify and retain only the significant interaction terms, streamlining the process and improving efficiency. This method is particularly advantageous for systems with many possible interactions but few dominant contributions, significantly improving efficiency over conventional methods. However, these advantages diminish in complex systems like ferroelectrics and multiferroics, where both the number of possible terms and significant interactions is large, leading to prohibitive computational costs.

In this work, we propose a novel fitting approach that combines Lasso regression with genetic algorithms, enabling the rapid construction of effective Hamiltonian models. In the following, we refer to this scheme as Lasso-GA Hybrid Method (LGHM). It is worth mentioning that this method can not only be used to construct effective Hamiltonian models for magnetic systems but also for Hamiltonian models that include atomic displacements. This is possible because we can express the energy as a linear combination of terms that include atomic displacements and strain components. [15,16] We benchmarked our method on a specific dataset and made applications on a model containing both magnetism and atomic displacements (monolayer $CrI_3$) and a magnetic system model (monolayer $Fe_3GaTe_2$), demonstrating its effectiveness. In the monolayer $Fe_3GaTe_2$, we further conducted an in-depth study of its higher-order interaction terms and their impact on the magnetic ground state and phase transition temperature, identifying the factors contributing to its high Curie temperature.

*Method*--Effective Hamiltonians are typically expressed as linear

combinations of specific basis functions. The general form of effective Hamiltonians can be written as:

$$H_{\text{eff}} = \sum_{j=1}^{p_{\max}} C_j h_j \qquad (1)$$

where $h_j$ are basis functions, $C_j$ are the corresponding coefficients and $p_{max}$ is the number of all possible terms. The basis functions can be obtained by Property analysis and simulation package for materials (PASP). [17] Determining the coefficients $C_j$ appears to be a straightforward linear regression problem, which is minimizing the sum of squares shown in equations (2), where $y_i$ is the exact value and $\widehat{y_i}$ is the value predicted by the linear model.

$$\sigma^2 = \frac{\sum_{i=1}^{n}(y_i - \widehat{y_i})^2}{n} \qquad (2)$$

However, there are two potential challenges when performing direct linear fitting. The first challenge is the limited number of data points. Linear regression requires the number of data points to exceed the number of parameters. Since the data points are obtained through DFT calculations, a large number of interaction terms would necessitate more DFT computations, resulting in higher computational costs. The second challenge is that even with sufficient data, too many interaction terms may lead to overfitting, reducing the model's predictive accuracy. To address the above challenges, we need a model that is both accurate and incorporates as few interaction terms as possible.

Our method combines Lasso regression [18,19] and genetic algorithm. [20] By defining the optimization problem and the genetic representation properly, the genetic algorithm can help identify the relevant interaction terms.

Inspired by MLMCH [7], we adopt the so-called $L_0$ regularization method. Similar to $L_1$ and $L_2$ regularization techniques commonly used in machine learning, we reformulate the linear model fitting task, which is minimizing the sum of squares as is shown before, into the minimization of the following equation (3), where $\lambda$ is a hyperparameter and $p$ is the number of selected terms. $\lambda^p$ is a penalty term. Minimizing $s$ allows us to balance both the accuracy of the model and the reduction in the number of model parameters.

$$s = \sigma^2 \times \lambda^p \tag{3}$$

In the genetic algorithm, we can represent the inclusion or exclusion of a specific interaction term using 1 and 0, respectively. The set of all selected terms in the model can thus be encoded as a binary gene consisting of 0s and 1s. The optimization problem is formulated as minimizing the $s$ mentioned above. However, in cases where the number of potential interaction terms reaches thousands, directly applying the genetic algorithm would be highly time-consuming. Therefore, our approach first employs Lasso regression, which can reduce small coefficients to zero [19], to pre-select relevant interaction terms before using the genetic algorithm.

The complete workflow of our LGHM is shown in Figure 1. First, we perform Lasso regression as a preliminary fitting and initial selection of interaction terms. Then, based on the selected terms, we proceed with further filtering using a genetic algorithm. Specifically, we initialize a set of genes, each representing whether a particular term is retained. We then perform linear regression on these genes and calculate their loss values. If the best gene satisfies certain conditions, such as no update in the best loss over 40 consecutive rounds, the loop exits; otherwise, the next evolutionary round begins. We select a subset of genes with relatively small loss values from all the results. These genes then undergo crossover and mutation to form new genes, and the process loops back to the linear regression and loss calculation step.

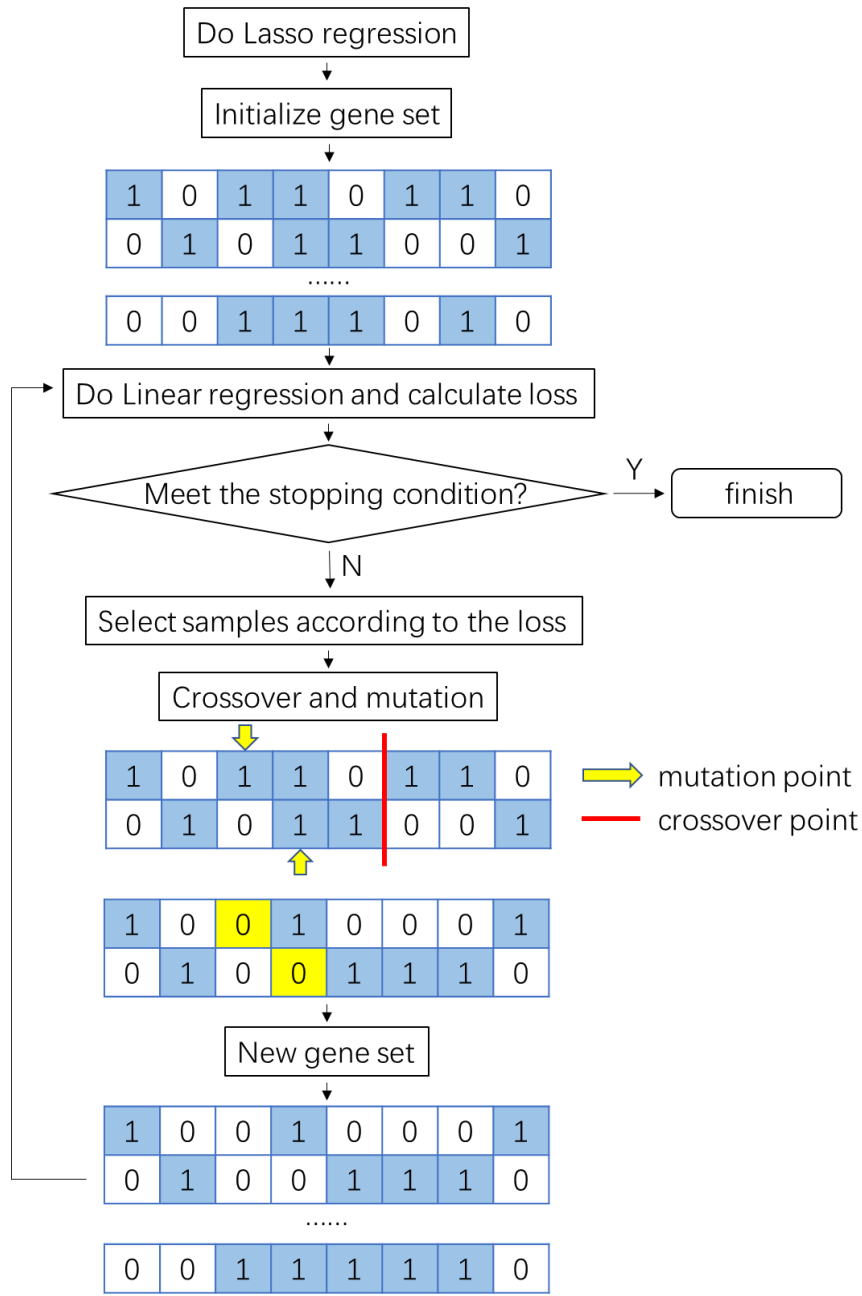

Figure 1. The flowchart of LGHM. The yellow arrows point out the mutation point and the red line means the crossover point.

*Benchmark*--We benchmark LGHM on a specific dataset designed to simulate the selection and fitting of relevant interaction terms from a pool of 6000 possible interactions. In this setup, 1000 interaction coefficients are randomly assigned non-zero values, with their magnitudes generated within a specified range. This range decreases progressively, simulating a realistic situation where nearest-neighbor, low-order interactions are larger, while higher-order, long-range interactions are smaller. This specific dataset contains a total of 5000 data points, with 4000 used for model fitting and the remaining 1000 reserved as the test set. LGHM takes 37604 seconds on two GPUs and 286910 seconds on CPU to get the fitting results of ten $\lambda$, while the previous method [7] fails to give the results even after running for two weeks.

The fitting results with $\lambda = 1.003$ are presented below. Our method produces a model with 668 interaction terms, achieving a mean absolute error (MAE) of 0.00212 and $R^2$ value of 0.9999. The parity plot of the model on the test dataset is shown in Figure 2. Among the coefficients we manually set, 601 terms have absolute values greater than 0.001, all of which are successfully identified by our method. In the supplemental materials [37], we provide the complete list of interaction terms with absolute values greater than 1.

This specific dataset demonstrates the capability of LGHM to efficiently identify significant interaction terms even when the total number of possible terms is extremely large. Additionally, it shows that the method remains effective even when the number of data points is smaller than the number of possible interaction terms.

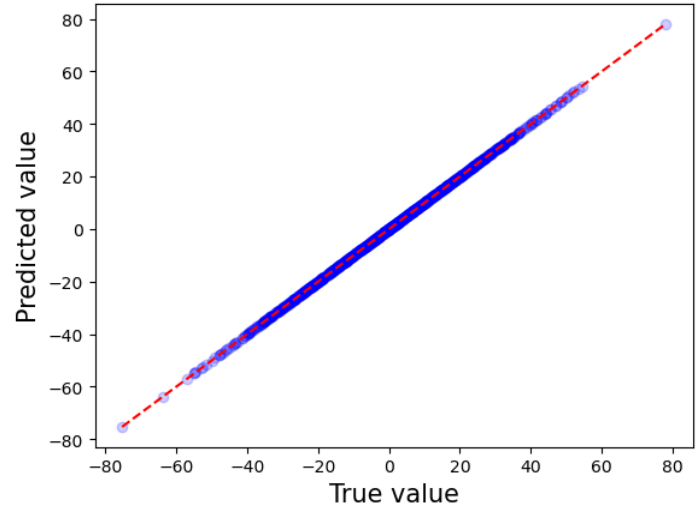

Figure 2. A plot of predicted values versus true values. Different blue dots indicate different samples in the test set.

*Application on monolayer* $CrI_3$--Significant progresses have recently been made in 2D magnets, making it possible to further push spintronic devices to the atomic level. [21,22] Among which, monolayer $CrI_3$ was reported as a 2D material with an out-of-plane ferromagnetic ground state. [23,24] This material exhibits a significant magnetic anisotropy, with a Curie temperature of 45K. [25] In order to get more insights into the microscopic magnetic properties of $CrI_3$, we develop its first-principles-based effective Hamiltonians. Here we adopt the symmetry-adapted cluster expansion method, as implemented in the PASP. [17] Such a method roots in cluster expansion that goes over all combinations of spin components and atomic displacements, enabling the generation of all possible forms of interactions, i.e., the invariants. The initial model contains enough invariants, including interactions up to the four-body and fourth-order (the truncation distances of the second, third and fourth order terms are set to

12 Bohr, 8 Bohr and 5 Bohr). After symmetry analyses, the number of possible interaction terms ($p_{max}$) is 25448. To fit and refine the coefficients of these invariants, a dataset of 5017 data points is generated using DFT calculations (the calculation details can be found in the supplemental materials [37]), where 4000 data points are used for fitting and 1017 data points are used for testing. In this dataset, each data point represents a structure consisting of 72 atoms, where both displacement degrees of freedom and magnetic state degrees of freedom are allowed.

Our method completes the fitting process of ten $\lambda$ in 79983 seconds on two GPUs and 327725 seconds on CPU. The result corresponding to the lowest mean squared error ($\lambda = 1.0009$) is selected. This yields a model with 737 interaction terms, achieving a mean absolute error (MAE) of 0.00235eV/u.c. and $R^2$ value of 0.9997. The parity plot of the model on the test dataset is shown in Figure 3b. Using the model obtained from our method, we perform Monte Carlo simulations and conjugate gradient (CG) optimizations to determine the ground state and phase transition temperature of monolayer $CrI_3$. As shown in Figure 3a, monolayer $CrI_3$ exhibits an out-of-plane ferromagnetic ground state with a phase transition temperature of 25K, in line with the previous studies. [23–25] The details of dataset and Monte Carlo simulations can be found in supplemental materials. [37]

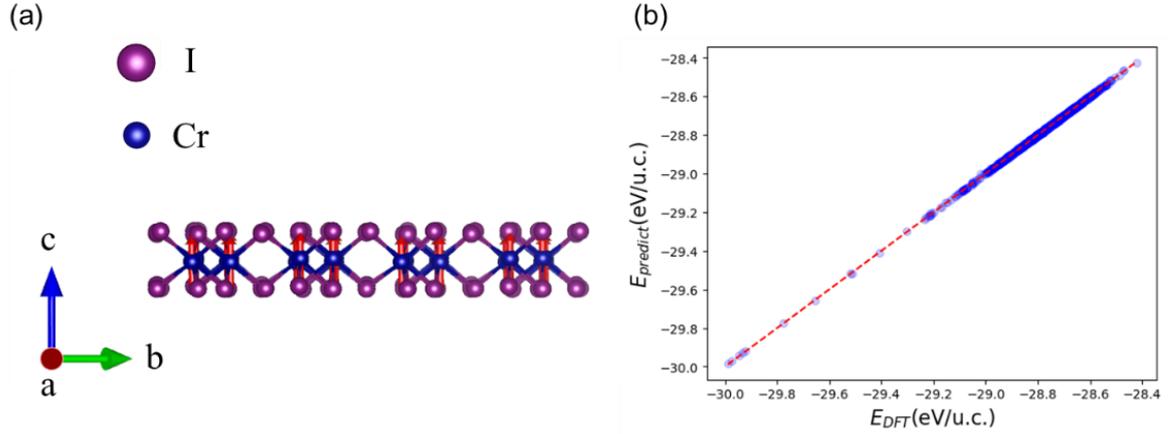

Figure 3. (a) The ground state of monolayer $CrI_3$. (b) A plot of predicted energies (per unit cell) versus computed values (using *ab initio* calculations)

*Application on monolayer $Fe_3GaTe_2$* -- $Fe_3GaTe_2$ (Figure 4) has emerged as a promising candidate material for next-generation spintronic devices due to its robust ferromagnetic properties down to the monolayer limit. Recent studies have highlighted its strong ferromagnetism, with notable findings indicating a Curie temperature ranging from 350 to 380 K. [26–29] Given these properties, our work aims to construct an exact spin Hamiltonian of $Fe_3GaTe_2$ and explore the microscopic origin of magnetic properties. To achieve this, we employed PASP software to generate all possible invariants. The truncation distances of the second, third and fourth order terms are set to 14 Bohr, 8.8 Bohr and 10 Bohr, which is up to the 10th, the 11th and the 21th nearest pairs, respectively. After symmetry analyses, the number of possible interaction terms ($p_{max}$) is 259. To fit and refine the coefficients of these invariants, a dataset of 1344 data points is generated using DFT calculations, where 1049 data points are used for fitting and 295 data points are used for testing.

It takes 774 seconds for our method on CPU to get the fitting results of ten $\lambda$. The result of the lowest mean squared error ($\lambda = 1.008$) is selected. Our method produces a model with 47 interaction terms, achieving a mean absolute error (MAE) of 0.00099eV/u.c. and $R^2$ value of 0.9998. Among the 47 interaction terms, the 14 most significant ones are listed in Table 1. We perform a MC simulation and CG optimization on our model to investigate the ground state and the phase transition on $Fe_3GaTe_2$. The simulation reveals the ground state is out-of-plane ferromagnetic, which well agrees with the experiment [26], with the Curie temperature being 460K (full model in Figure 5). The details of dataset and Monte Carlo simulations can be found in supplemental materials. [37]

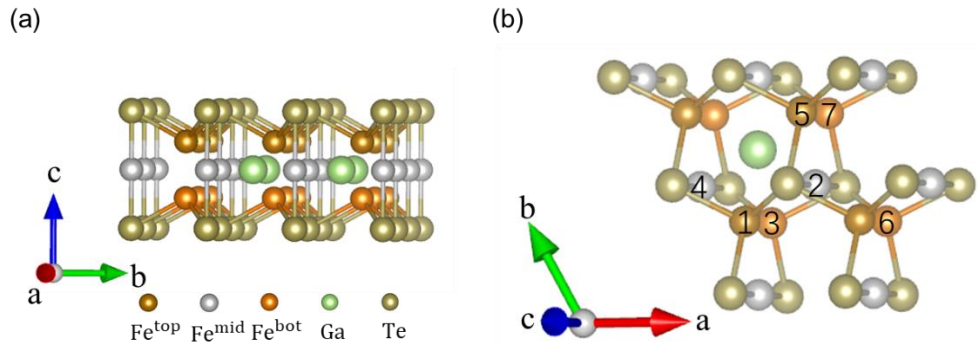

Figure 4. (a) The structure of monolayer $Fe_3GaTe_2$. (b) The numbering of the Fe atoms corresponds to the subscripts of the magnetic moments in Table 1.

Although the ground state of 2D $Fe_3GaTe_2$ is out-of-plane ferromagnetic, as indicated by both experimental results and our simulations, its microscopic mechanism remains unclear. According to our calculations, as shown in Table 1, the nearest-neighbor Heisenberg interaction exhibits an antiferromagnetic tendency. This contrasts to previous studies [27,30], whose results suggest that the nearest-neighbor Heisenberg interaction is ferromagnetic. This discrepancy may arise from differences in the employed models. In our study, numerous high-order interactions are considered, particularly the biquadratic interactions, which favor the ferromagnetic state and has a non-negligible magnitude.

To illustrate the significance of those interaction terms, MC simulations and CG optimizations were performed on two additional model: model A, incorporating the fourteen interaction terms listed in table 1 and model B, considering only the single ion anisotropy and exchange coupling mentioned in table 1. Both simulations reveal an out-of-plane ferromagnetic ground state. Model A indicates a Curie temperature of 350K, which is close to that of the full model, while model B yields a Curie temperature of 120K. The significant difference in Curie temperatures between these models implies the crucial role of higher-order interactions. This is understandable since the magnitudes of higher-order interactions are comparable to those of Heisenberg exchange interactions, and their cooperative effect tends to favor the ferromagnetic state. Table 2 presents the energy disparities between different magnetic states calculated using DFT, the full model and the model with selected terms, further demonstrating the energy contribution of high order interactions. In contrast to Fe3GeTe2 [3,37], whose Curie temperature primarily originates from strong Heisenberg exchange interactions, Fe3GaTe2 exhibits a higher Curie temperature, which can be mainly attributed to its fourth-order interactions. This finding significantly deepens the understanding of high-order interaction-driven high Curie temperatures and highlights the potential of Fe3GaTe2 in advanced spintronic applications.

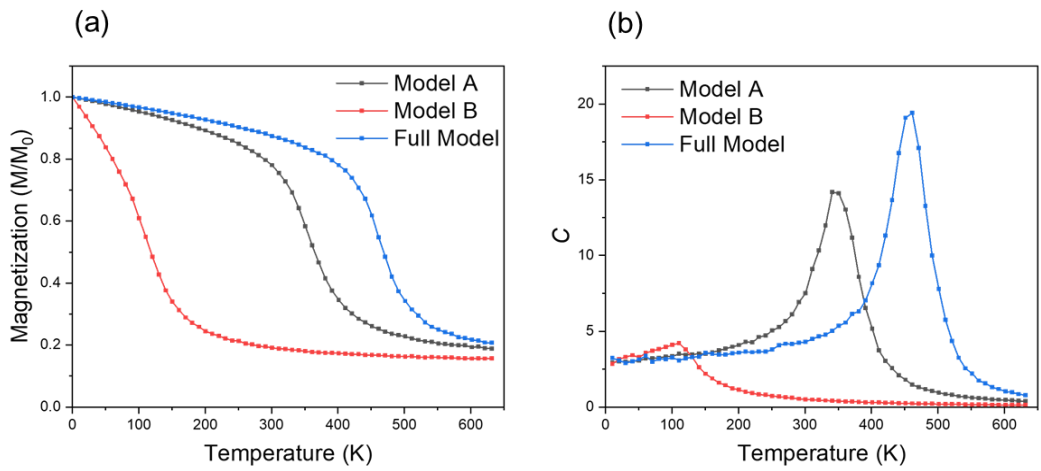

Figure 5. (a) The average Magnetization of Fe atoms in different temperatures of monolayer $Fe_3GaTe_2$. Lines in different color are using different models. Model A contains only the A and J interactions listed in table 1, Model B contains all the interactions listed in table 1, while the Full Model refers to the model with all the 47 interactions find by our method.

Table 1. selected magnetic parameters of monolayer $Fe_3GaTe_2$. The superscripts 't', 'b', and 'm' represent the Fe atoms located in the top layer, bottom layer, and middle layer, which is illustrated in figure 4(a) respectively. *A* stands for single ion anisotropy, *J* stands for exchange

couplings, *D* stands for DMI, *B* stands for biquadratic interaction and *K* stands for fourth-order coefficients. The subscripts '1', '2' refer to the atoms indicated in figure 4(b).

| Index | symbol | value | invariant | Index | symbol | value | Invariant |
|---|---|---|---|---|---|---|---|
| 1 | $A_{zz}^t$ | -2.0 | $S_z^2$ | 8 | $B^{tb}$ | -7.4 | $(\vec{S_1} \cdot \vec{S_3})^2$ |
| 2 | $A_{zz}^m$ | 1.4 | $S_z^2$ | 9 | $K_1$ | -5.7 | $(\vec{S_3} \cdot \vec{S_1}) \cdot (\vec{S_2} \cdot \vec{S_3})$ |
| 3 | $J^{tb1}$ | 6.4 | $\vec{S_1} \cdot \vec{S_3}$ | 10 | $K_2$ | -5.7 | $(\vec{S_3} \cdot \vec{S_1}) \cdot (\vec{S_2} \cdot \vec{S_1})$ |
| 4 | $J^{tm}$ | -11.2 | $\vec{S_1} \cdot \vec{S_2}$ | 11 | $K_3$ | -5.1 | $(\vec{S_1} \cdot \vec{S_2}) \cdot (\vec{S_2} \cdot \vec{S_3})$ |
| 5 | $J^{tt}$ | 4.1 | $\vec{S_1} \cdot \vec{S_5}$ | 12 | $K_4$ | -5.3 | $(\vec{S_2} \cdot \vec{S_3}) \cdot (\vec{S_2} \cdot \vec{S_6})$ |
| 6 | $D^{tt}$ | -5.3 | $S_{1x}S_{5y} - S_{5x}S_{1y}$ | 13 | $K_5$ | -6.3 | $(\vec{S_2} \cdot \vec{S_3}) \cdot (\vec{S_2} \cdot \vec{S_5})$ |
| 7 | $J^{tb2}$ | -6.4 | $\vec{S_3} \cdot \vec{S_5}$ | 14 | $K_6$ | -4.6 | $(\vec{S_3} \cdot \vec{S_7}) \cdot (\vec{S_1} \cdot \vec{S_5})$ |

Table 2 The energy difference per atom between different magnetic states of monolayer FGT in units of meV/atom. OP and IP stand for out of plane and in plane, while FM and FIM stand for ferromagnetic and ferrimagnetic.

Special stands for a specific spin state, which is shown in the supplemental material. [37] Full model refers to the model with all the interactions find by LGHM. Model A refers to the model with only the A and J interactions listed in table 1. Model B refers to the model with the interaction listed in table 1. The spin configurations of the magnetic states are shown in supplemental materials. [37]

| Method \ Magnetic state | OP-FM | IP-FM | OP-FIM | special |
|---|---|---|---|---|
| DFT | 0 | 0.15 | 66.79 | 42.54 |
| Full model | 0 | 0.15 | 66.69 | 41.44 |
| Model A | 0 | 0.43 | 21.84 | 11.25 |
| Model B | 0 | 0.43 | 54.55 | 35.19 |

In conclusion, we propose the LGHM for obtaining effective Hamiltonian models. We benchmark this method on a specific dataset, demonstrating its effectiveness. Subsequently, we apply our approach to monolayer $CrI_3$ and $Fe_3GaTe_2$, where we identify the interaction terms and performed Monte Carlo simulations using these terms. The results yield reasonable magnetic ground states and Curie temperatures, further validating the robustness of our method. Notably, in the case of monolayer $Fe_3GaTe_2$, we find that the high-order interactions are crucial for its high Curie temperature.

Compared to the previous MLMCH method, our model can obtain results more efficiently when dealing with a large number of possible interaction terms, for the sake of the initial selection performed using L1 regularization before genetic algorithm. This is particularly common when the model considers atomic displacements, making LGHM especially suitable for such complex scenarios.

We acknowledge financial support from NSFC (Grants No. 12188101, No. 12174060, and No. 12274082, the Guangdong Major Project of the Basic and Applied basic Research (Future functional materials under extreme conditions C2021B0301030005), Shanghai Pilot Program for Basic Research-FuDan University 21TO1400100 (23T0017), and Shanghai Science and Technology Program (23JC1400900). C.X. also acknowledges support from the Shanghai Science and Technology Committee (Grant No. 23ZR1406600), Innovation Program for Quantum Science and Technology (2024ZD0300102) and the Xiaomi Young Talents Program. B.Z. also acknowledges the support from the China Postdoctoral Science Foundation (Grant No.2022M720816, 2024T170152).

**Reference**

[1] W. Kohn and L. J. Sham, Self-Consistent Equations Including Exchange and Correlation Effects, Phys. Rev. **140**, A1133 (1965).

[2] G. Kresse and J. Furthmüller, Efficient iterative schemes for ab initio total-energy calculations using a plane-wave basis set, Phys. Rev. B **54**, 11169 (1996).

[3] C. Xu, X. Li, P. Chen, Y. Zhang, H. Xiang, and L. Bellaiche, Assembling Diverse Skyrmionic Phases in $Fe_3GeTe_2$ Monolayers, Advanced Materials **34**, 2107779 (2022).

[4] X. Li, C. Xu, B. Liu, X. Li, L. Bellaiche, and H. Xiang, Realistic Spin Model for Multiferroic $NiI_2$, Phys. Rev. Lett. **131**, 036701 (2023).

[5] S. Picozzi and C. Ederer, First Principles Studies of Multiferroic Materials, J. Phys.: Condens. Matter **21**, 303201 (2009).

[6] T. Moriya, Anisotropic Superexchange Interaction and Weak Ferromagnetism, Phys. Rev. **120**, 91 (1960).

[7] X.-Y. Li, F. Lou, X.-G. Gong, and H. Xiang, Constructing realistic effective spin Hamiltonians with machine learning approaches, New J. Phys. **22**, 053036 (2020).

[8] F. Illas, I. P. R. Moreira, C. de Graaf, and V. Barone, Magnetic coupling in biradicals, binuclear complexes and wide-gap insulators: a survey of ab initio wave function and density functional theory approaches, Theor Chem Acc **104**, 265 (2000).

[9] P. Reinhardt, M. P. Habas, R. Dovesi, I. de P. R. Moreira, and F. Illas, Magnetic coupling in the weak ferromagnet $CuF_2$, Phys. Rev. B **59**, 1016 (1999).

[10] H. Xiang, C. Lee, H.-J. Koo, X. Gong, and M.-H. Whangbo, Magnetic properties and energy-mapping analysis, Dalton Trans. **42**, 823 (2012).

[11] H. J. Xiang, E. J. Kan, S.-H. Wei, M.-H. Whangbo, and X. G. Gong, Predicting the spin-lattice order of frustrated systems from first principles, Phys. Rev. B **84**, 224429 (2011).

[12] N. S. Fedorova, C. Ederer, N. A. Spaldin, and A. Scaramucci, Biquadratic and ring exchange interactions in orthorhombic perovskite manganites, Phys. Rev. B **91**, (2015).

[13] P. Novák, I. Chaplygin, G. Seifert, S. Gemming, and R. Laskowski, Ab-initio calculation of exchange interactions in $YMnO_3$, Computational Materials Science **44**, 79 (2008).

[14] H.-F. Zhu, H.-Y. Cao, Y. Xie, Y.-S. Hou, S. Chen, H. Xiang, and X.-G. Gong, Giant biquadratic interaction-induced magnetic anisotropy in the iron-based superconductor $A_xFe_{2-y}Se_2$, Phys. Rev. B **93**, 024511 (2016).

[15] W. Zhong, D. Vanderbilt, and K. M. Rabe, First-principles theory of ferroelectric phase transitions for perovskites: The case of $BaTiO_3$, Phys. Rev. B **52**, 6301 (1995).

[16] L. Bellaiche, A. García, and D. Vanderbilt, Finite-Temperature Properties of Pb($Zr_{1-x}Ti_x$)$O_3$ Alloys from First Principles, Phys. Rev. Lett. **84**, 5427 (2000).

[17] F. Lou, X. Y. Li, J. Y. Ji, H. Y. Yu, J. S. Feng, X. G. Gong, and H. J. Xiang, PASP: Property analysis and simulation package for materials, The Journal of Chemical Physics **154**, 114103 (2021).

[18] R. Tibshirani, Regression Shrinkage and Selection Via the Lasso, Journal of the Royal Statistical Society Series B: Statistical Methodology **58**, 267 (1996).

[19] L. Freijeiro-González, M. Febrero-Bande, and W. González-Manteiga, A Critical Review of LASSO and Its Derivatives for Variable Selection Under Dependence Among Covariates, International Statistical Review **90**, 118 (2022).

[20] J. H. Holland, Genetic Algorithms, SCIENTIFIC AMERICAN (1992).

[21] C. Gong et al., Discovery of intrinsic ferromagnetism in two-dimensional van der Waals crystals, Nature **546**, 265 (2017).

[22] M. Bonilla, S. Kolekar, Y. Ma, H. C. Diaz, V. Kalappattil, R. Das, T. Eggers, H. R. Gutierrez, M.-H. Phan, and M. Batzill, Strong room-temperature ferromagnetism in $VSe_2$ monolayers on van der Waals substrates, Nature Nanotech **13**, 289 (2018).

[23] B. Huang et al., Layer-dependent ferromagnetism in a van der

Waals crystal down to the monolayer limit, Nature **546**, 270 (2017).

[24]  J. L. Lado and J. Fernández-Rossier, On the origin of magnetic anisotropy in two dimensional $CrI_3$, 2D Mater. **4**, 035002 (2017).

[25]  J. F. Dillon and C. E. Olson, Magnetization, Resonance, and Optical Properties of the Ferromagnet $CrI_3$, Journal of Applied Physics **36**, 1259 (1965).

[26]  G. Zhang, F. Guo, H. Wu, X. Wen, L. Yang, W. Jin, W. Zhang, and H. Chang, Above-room-temperature strong intrinsic ferromagnetism in 2D van der Waals $Fe_3GaTe_2$ with large perpendicular magnetic anisotropy, Nat Commun **13**, 1 (2022).

[27]  J.-E. Lee, S. Yan, S. Oh, J. Hwang, J. D. Denlinger, C. Hwang, H. Lei, S.-K. Mo, S. Y. Park, and H. Ryu, Electronic Structure of Above-Room-Temperature van der Waals Ferromagnet $Fe_3GaTe_2$, Nano Lett. **23**, 11526 (2023).

[28]  G. Hu et al., Room-Temperature Antisymmetric Magnetoresistance in van der Waals Ferromagnet $Fe_3GaTe_2$ Nanosheets, Advanced Materials **36**, 2403154 (2024).

[29]  S. Wu et al., Robust ferromagnetism in wafer-scale $Fe_3GaTe_2$ above room-temperature, Nat Commun **15**, 10765 (2024).

[30]  X. Li, M. Zhu, Y. Wang, F. Zheng, J. Dong, Y. Zhou, L. You, and J. Zhang, Tremendous tunneling magnetoresistance effects based on van der Waals room-temperature ferromagnet $Fe_3GaTe_2$ with highly spin-

polarized Fermi surfaces, Applied Physics Letters **122**, 082404 (2023).

[31]   F. Pedregosa et al., Scikit-learn: Machine Learning in Python, Journal of Machine Learning Research, Vol 12, pp. 2825-2830 (2011)

[32]   J. P. Perdew, K. Burke, and M. Ernzerhof, Generalized Gradient Approximation Made Simple, Phys. Rev. Lett. **77**, 3865 (1996).

[33]   P.-W. Ma and S. L. Dudarev, Constrained density functional for noncollinear magnetism, Phys. Rev. B **91**, 054420 (2015).

[34]   S. L. Dudarev, G. A. Botton, S. Y. Savrasov, C. J. Humphreys, and A. P. Sutton, Electron-energy-loss spectra and the structural stability of nickel oxide: An LSDA+U study, Phys. Rev. B **57**, 1505 (1998).

[35]   U. H. E. Hansmann, Parallel tempering algorithm for conformational studies of biological molecules, Chemical Physics Letters **281**, 140 (1997).

[36]   M. R. Hestenes and E. Stiefel, Methods of Conjugate Gradients for Solving Linear Systems, Journal of research of the National Bureau of Standards, **49**, 409 (1952).

[37]   Z.-X. Shen, X. Bo, K. Cao, X. Wan, and L. He, Magnetic ground state and electron-doping tuning of Curie temperature in Fe 3 GeTe 2 : First-principles studies, Phys. Rev. B **103**, 085102 (2021).

[38] See Supplemental Material at http://link for more information about details of dataset and Monte Carlo simulations, which includes Refs. [2, 18, 19, 20, 31-36]